# High impedance TES bolometers for EDELWEISS

S. Marnieros[1] • E. Armengaud[2] • Q. Arnaud[3] • C. Augier[3] • A. Benoît[4] • L. Bergé[1] • J. Billard[3] • A. Broniatowski[1] • P. Camus[4] • A. Cazes[3] • M. Chapellier[1] • F. Charlieux[3] • M. De Jésus[3] • L. Dumoulin[1] • K. Eitel[5] • J.-B. Fillipini[3] • D. Filosofov[6] • J. Gascon[3] • A. Giuliani[1] • M. Gros[2] • Y. Jin[7] • A. Juillard[3] • M. Kleifges[8] • H. Lattaud[3] • D. Misiak[3] • X.-F. Navick[2] • C. Nones[2] • E. Olivieri[1] • C. Oriol[1] • P. Pari[9] • B. Paul[2] • D. Poda[1] • S. Rozov[6] • T. Salagnac[3] • V. Sanglard[3] • L. Vagneron[3] • E. Yakushev[6] • A. Zolotarova[1]

[1] *Université Paris-Saclay, CNRS/IN2P3, IJCLab, 91405 Orsay, France*
[2] *IRFU, CEA, Université Paris-Saclay, F-91191 Gif-sur-Yvette, France*
[3] *Univ Lyon, Université Lyon 1, CNRS/IN2P3, IP2I-Lyon, F-69622 Villeurbanne, France*
[4] *Institut Néel, CNRS/UJF, 38042 Grenoble, France*
[5] *Karlsruher Institut für Technologie, Institut für Astroteilchenphysik, Postfach 3640, 76021 Karlsruhe, Germany*
[6] *JINR, Laboratory of Nuclear Problems, Joliot-Curie 6, 141980 Dubna, Moscow Region, Russian Federation*
[7] *C2N, CNRS, Université Paris-Sud, Université Paris-Saclay, 91120 Palaiseau, France*
[8] *Karlsruher Institut für Technologie, Institut für Prozessdatenverarbeitung und Elektronik, Postfach 3640, 76021 Karlsruhe, Germany*
[9] *IRAMIS, CEA, Université Paris-Saclay, F-91191 Gif-sur-Yvette, France*

**Abstract** The EDELWEISS collaboration aims for direct detection of light dark matter using germanium cryogenic detectors with low threshold phonon sensor technologies and efficient charge readout designs. We describe here the development of Ge bolometers equipped with high impedance thermistors based on a $Nb_xSi_{1-x}$ TES alloy. High aspect ratio spiral designs allow the TES impedance to match with JFET or HEMT front-end amplifiers. We detail the behavior of the superconducting transition properties of these sensors and the detector optimization in terms of sensitivity to out-of-equilibrium phonons. We report preliminary results of a 200 g Ge detector that was calibrated using $^{71}$Ge activation by neutrons at the LSM underground laboratory.

**1 Introduction**

Development of very low threshold detectors is in the heart of several experiments focused on light dark matter search or detection of coherent scattering of neutrinos by nuclei. Low temperature bolometers using transition edge sensors (TES) are among the most promising devices in these domains and are developed by several groups worldwide. In the following sections we detail the fabrication and optimization of Ge bolometers equipped with a $Nb_xSi_{1-x}$ TES that is tailored to JFET or HEMT based preamplifiers, and do not require a SQUID read-out. Preliminary tests on 200 g Ge prototypes are very promising for the EDELWEISS-SubGeV dark matter program [1].

**2 $Nb_xSi_{1-x}$ superconducting transition properties**

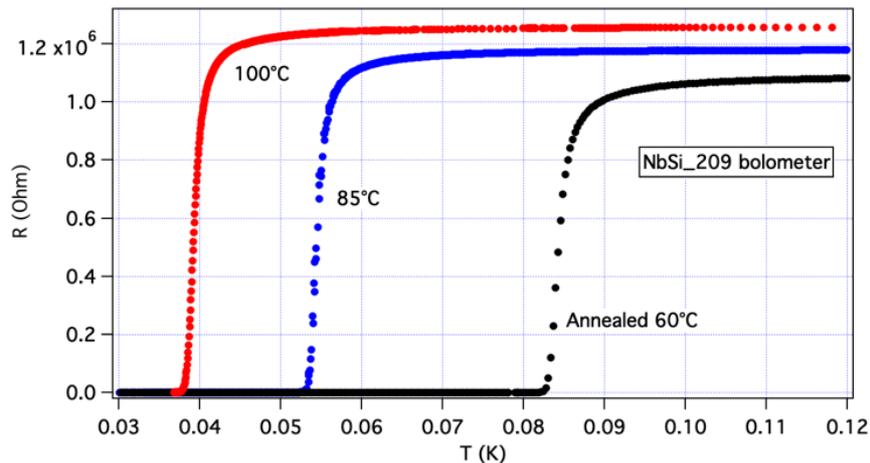

**Fig. 1** Annealing effect of a 100 nm thick $Nb_{0,132}Si_{0,868}$ film deposited on a 200 g Ge crystal. The sample has been annealed three times during one hour at increasing temperatures. After each annealing, the superconducting transition curve is measured in a dilution refrigerator. This method allows to lower the TES critical temperature in a reproducible way. (Color figure online)

$Nb_xSi_{1-x}$ films are fabricated at IJCLab by electron-beam co-evaporation of Nb and Si crucibles, into a high vacuum chamber equipped with an Inficon-IC6 deposition controller that allows precise adjustment of the sample thickness and composition. This fabrication process provides amorphous

# High impedance TES bolometers for EDELWEISS

$Nb_xSi_{1-x}$ thin layers exhibiting a normal-to-superconducting transition for stochiometric concentrations of Nb above approximately 13%. Increasing the Nb concentration allows to precisely adjust the film critical temperature between zero and several degrees Kelvin [2]. An additional fine-tuning of the critical temperature is also possible after the film fabrication by annealing of the sample between 50 °C and 350 °C (Fig. 1). $Nb_xSi_{1-x}$ films can be deposited on any type of substrate at room temperature, and are compatible with standard micro-lithography processing techniques. To ensure optimal homogeneity of the $Nb_xSi_{1-x}$ composition and film properties, the substrate is rotating during the co-evaporation process.

An interesting aspect of thin $Nb_xSi_{1-x}$ layers is related to their high normal-state resistivity, that is two to three orders of magnitude higher than usual TES films like W, Ti, Al or proximity effect bi-layers [2]. This property can be combined to a specific $Nb_xSi_{1-x}$ design in order to match its noise to either SQUID based electronics [3] or high input-impedance JFET and HEMT preamplifiers [4]. In this paper we will focus on the latter solution, requiring a $Nb_xSi_{1-x}$ TES impedance of the order of 1 MΩ, in order to bring the detector noise above the read-out amplifier noise. This can be achieved by patterning the TES to a high aspect ratio spiral or meander, with a total line-length to line-width of the order of $10^3$ to $10^4$. The $Nb_xSi_{1-x}$ volume can be tailored upon our needs by adjusting the line-width in the range of 1 to 200 μm while keeping a constant aspect ratio.

It is important to notice that the high aspect-ratio geometry is not compatible with a strong negative electrothermal feedback of the TES, associated to a constant voltage biasing scheme. Negative electrothermal feedback is interesting because it induces temperature self-regulation of the sensor, faster response time and lower Johnson noise. In a constant voltage-biased high aspect ratio TES it is not possible to achieve a homogeneous voltage gradient through the sensor. There is a bias limit above which the TES line will split in superconducting domains with zero voltage drop and normal state domains with finite voltage drop, affecting substantially the detector sensitivity and resolution. Nevertheless, high aspect ratio $Nb_xSi_{1-x}$ TES can perform conveniently if current-biased below the bias instability limit and achieve high sensitivity and thermal-noise limited operation [4].

The maximum current that can be applied to a high aspect ratio $Nb_xSi_{1-x}$ layer results from the competition between the cooling of the TES electrons via the electron-phonon coupling and the positive electrothermal effect related to Joule heating variations in the sensor. The TES becomes instable when a small temperature fluctuation results to higher Joule dissipation than the electron-phonon coupling can evacuate:

$$\Delta P_{electrothermal} > \Delta P_{electron-phonon} \Rightarrow I^2 \Delta R > G_{e-ph} \Delta T \quad (1)$$

$I$ is the TES current bias, $G_{e-ph}$ the electron-phonon coupling constant of the $Nb_xSi_{1-x}$ and $\Delta R/\Delta T$ the sensor transition slope measured at low bias.

The maximum current ($I_{max}$) that can flow through the $Nb_xSi_{1-x}$ line before emergence of instability effects is given by equation (2).

$$I_{max} = \sqrt{\frac{G_{e-ph}}{\frac{\Delta R}{\Delta T}}} \quad (2)$$

The $Nb_xSi_{1-x}$ electron-phonon coupling is described by equation 3 [4].

$$G_{e-ph} = 10^9 \, \Omega_{NbSi} \, T^4 \quad (3)$$

$\Omega_{NbSi}$ is the volume of the $Nb_xSi_{1-x}$ film in units of $m^3$.

So, the maximum current bias is proportional to $T^2$ and will limit the sensitivity of $Nb_xSi_{1-x}$ TES detectors at low temperature.

## 3 $Nb_xSi_{1-x}$ TES detectors

An interesting aspect of TES detectors based on $Nb_xSi_{1-x}$ layers is their sensitivity to out-of-equilibrium phonons. During a transient regime following a particle or photon interaction with a massive bolometer, out-of-equilibrium phonons are created and propagate the deposited energy through the detector crystal. These phonons interact strongly with the electrons of the TES, releasing their energy and significantly heating the sensor. In an optimally designed device, the temperature of the sensor can rise well above that of the crystal. Sensitivity to out-of-equilibrium phonons has already been reported on several TES bolometers [5,6], allowing to enhance their performances in terms of energy resolution and signal-to-noise ratio. In the case of a $Nb_xSi_{1-x}$ Ge bolometer operating below 50 mK, the thermal relaxation time between the sensor and the Ge crystal becomes much slower (> 1 ms) than the out-of-equilibrium phonon life-time (typically 200 μs in a 200 g Ge bolometer). The TES layer will integrate the energy of incoming phonons, and undergo a temperature rise of the order of:

$$\Delta T_{NbSi} = \varepsilon \frac{E}{C_{NbSi}} \quad (4)$$

$\varepsilon E$ is the part of the deposited energy integrated by the TES electron-bath. $C_{NbSi}$ is the $Nb_xSi_{1-x}$ heat capacity.

The detector sensitivity is proportional to the fraction $\varepsilon$ of the deposited energy that is transferred to the TES through out-of-equilibrium phonons. To maximize this former it is crucial to carefully design the detector in order to reduce trapping of high-energy phonons on the crystal surface and in deposited layers other than the TES itself. Specific "phonon-trap" layers can be coupled to the TES [5,6] and increase detector sensitivity at the expense of a more complicated fabrication and optimization process. They are considered for a future upgrade of our detectors. In the case of Ge bolometers with

# High impedance TES bolometers for EDELWEISS

simultaneous heat and ionization read-out, the presence of large charge collecting electrodes induce very efficient down-conversion of high energy phonons and will considerably affect the sensitivity of the $Nb_xSi_{1-x}$ sensor. In order to minimize this effect, we have studied several design solutions on a 35 g Ge detector. This latter was based on a high purity ($N_A-N_D < 2\ 10^{10}$ cm$^{-3}$) n-type Ge crystal, manufactured in a cylindrical shape (diameter 20 mm, height 20 mm). A 50 nm thick and 6 mm diameter spiral-shaped $Nb_{0.132}Si_{868}$ sensor was partially covering the top surface of the crystal, while the bottom one was processed in seven different ways (Table 1).

| Cryogenic run | Process on bottom surface | Sensitivity (nV/keV) |
|---|---|---|
| 1 | Bare Ge surface | 1400 |
| 2 | a-Ge:H + Al (200 nm, full coverage) | 280 |
| 3 | a-Ge:H (no electrode) | 1400 |
| 4 | a-Ge:H + Al (20 nm, full coverage) | 280 |
| 5 | a-Ge:H + Al grid (2% coverage) | 800 |
| 6 | a-Ge:H + Nb (50nm, full coverage) | 370 |
| 7 | a-Ge:H + Nb + XeF$_2$ on lateral side | < 100 |

**Table 1.** Detector sensitivity for different configurations of the Ge bottom surface. Aluminum wet-etching was used to completely remove the Al layer after runs 2 and 5, and to pattern the 20 nm Al layer to a grid after run 4.

For each of the cryogenic runs, the sensitivity of the detector was measured under the same TES current bias and operating temperature, using the 60 keV line of an $^{241}$Am source. Adding a superconducting Al or Nb layer on the bottom Ge surface results to an important sensitivity drop due to an efficient down-conversion of high-energy phonons. The Al layer has a higher down-conversion efficiency due to its lower critical temperature and much longer quasiparticle time-life. We observe no influence of the Al film thickness on the detector sensitivity for 20 and 200 nm layers. However, replacing a full Al coating by a 2% coverage Al grid results to a substantial sensitivity gain and seems to be a good compromise for our detectors. Use of a Nb grid could be even more advantageous but is more difficult to process. An amorphous 30 nm thick Ge layer, passivated by H, is often deposited below the charge collecting electrodes to reduce the detector leakage current and does not seem to affect the TES sensitivity in our case (cryogenic run 3). In the last run we tested a XeF$_2$ dry-etching of the Ge lateral surface that is also a solution for reducing the leakage currents [7]. This etching produces a very rough Ge surface, resulting to an important phonon-scattering process that affects dramatically the $Nb_xSi_{1-x}$ sensitivity.

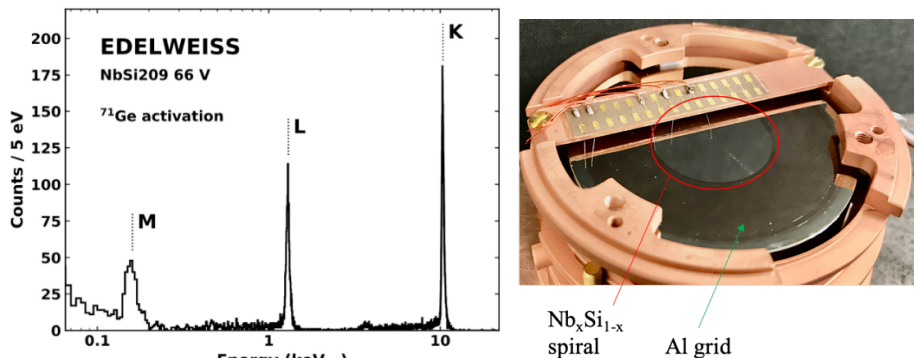

**Fig. 2.** *Left:* calibration of a 200 g Nb$_x$Si$_{1-x}$ Ge bolometer using $^{71}$Ge activation by neutrons. The Ge crystal was biased at 66 V for NTL amplification. *Right:* the top surface of the detector is covered by a TES layer and a 2 % coverage Al grid. The bottom Ge surface is fully covered by an Al grid. (Color online)

Following these results, we performed an upscale to 200 g Ge detectors equipped with a 20 mm diameter Nb$_x$Si$_{1-x}$ spiral TES, for the EDELWEISS dark matter experiment. Operating these detectors in the so called Neganov-Trofimov-Luke (NTL) amplification mode [1 and references therein] resulted in a baseline energy resolution of approximately 5 eV$_{ee}$ rms (Fig. 2). Dark matter constrains obtained from these detectors at the Laboratoire Souterrain de Modane will be reported on a separate paper. The EDELWEISS collaboration is pursuing a development program to achieve single-electron threshold Ge bolometers coupled to efficient background rejection techniques.